\documentclass[aps,prb,twocolumn,groupedaddress]{revtex4}
\usepackage{graphicx,latexsym,lscape,longtable}

\begin{document}

\title{Superconductivity and short range order in metallic glasses Fe$_{x}$Ni$_{1-x}$Zr$_{2}$}

\author{J.\ Lefebvre, M.\ Hilke, and Z.\ Altounian}

\affiliation{ Department of Physics, McGill University, Montr\'eal, Canada
H3A 2T8.}

\begin{abstract}

In amorphous superconductors, superconducting and vortex pinning properties are strongly linked to the absence of long range order. Consequently, superconductivity and vortex phases can be studied to probe the underlying microstructure and order of the material. This is done here from resistance and local magnetization measurements in the superconducting state of Fe$_{x}$Ni$_{1-x}$Zr$_{2}$ metallic glasses with $0\leq x \leq 0.6$. Firstly, we present typical superconducting properties such as the critical temperature and fields and their dependence on Fe content in these alloys. Then, the observations of peculiar clockwise hysteresis loops, wide double-step transitions and large magnetization fluctuations in glasses containing a large amount of Fe are analyzed to reveal a change in short range order with Fe content. 

\end{abstract}
\maketitle
\section{Introduction}
Since the synthesis of the first amorphous alloy superconductors by vapor deposition by B\"{u}ckel and Hilsch \cite{BuckelZphys138} in 1954, and the following, fabricated in the mid-1970s by electron beam evaporation by Collver and Hammond \cite{CollverPRL30}, our understanding of superconducting phenomenon in this new class of superconductors has greatly evolved. Although these first attempts at making amorphous alloys were plagued by the inconvenient instability of the amorphous phase at room temperature, they have served to expose the differences between superconductivity in amorphous materials and in their crystalline counterpart. Nowadays, several techniques based on the rapid cooling of the melt are used to fabricate stable amorphous alloys and the past 25 years have seen the publication of many studies about superconductivity in such metallic glasses, especially those composed of transition metal alloys \cite{AltounianPRB27, Sabouri, Trudeau, Ristic, Hamed, Flodin, KarkutPRB28}. These studies have, among other things, discussed the importance of including effects due to spin fluctuations in the predictions of T$_{c}$, especially in alloys containing Ni, Co, or Fe. Additionally, they have identified the important role played by microstructure and short range order (SRO) in amorphous materials. Since microstructure critically depends on the fabrication process and the melt cooling rate, values for different superconducting characteristics such as the critical temperature T$_{c}$ from different labs for the same alloy composition often vary significantly.\\
Vortices in type II superconductors can typically also reveal important information about material structure. In particular, defects and dislocations provide pinning sites which prevent vortex movement and enhance pinning. On the contrary, the absence of long range order in amorphous materials greatly decreases pinning properties. Moreover, several superconducting properties, for instance the shape and width of the B$_{c2}$ transition, or magnetization hysteresis loops are direct consequences of the homogeneity of the material and its flux trapping capacities. Binary and pseudo-binary compounds composed of the early transition (ET) metal Zr and late transition (LT) metal Cu, Ni, Co and Fe in the form LT$_{x}$ET$_{1-x}$ and (LT$_{x}^{a}$LT$_{1-x}^{b})_{y}$ET$_{1-y}$ have shown excellent glass-forming abilities over a wide compositional range \cite{AltounianPRB27, AltounianPRB49, DikeakosJNCS250, Hamed, KarkutPRB28}. In this article, we study such a pseudo-binary compound, namely Fe$_{x}$Ni$_{1-x}$Zr$_{2}$ superconducting glasses with $0\leq x\leq0.6$, based on electric transport and local magnetization measurements. The high purity and the amorphous nature of these alloys, conferring them extremely weak pinning properties, have previously allowed us to investigate transversely ordered dynamic vortex phases \cite{HilkePRL91, LefebvrePRB74, LefebvrePRB78}. Here, we exploit the relationship between vortex pinning and material structure to reveal a change in SRO in this glass series. This conclusion is reached based on the observation of peculiar hysteresis loops and magnetization fluctuations in alloys containing a relatively large amount of Fe. Superconducting properties of the metallic glasses and their dependence on the Fe content $x$ are also presented.

\section{Experimental methods}

\begin{figure}
[ptbh]
\begin{center}
\includegraphics[
trim=0.000000in 3.807070in 0.000000in 0.000000in,
height=2.7155in,
width=3.2214in
]%
{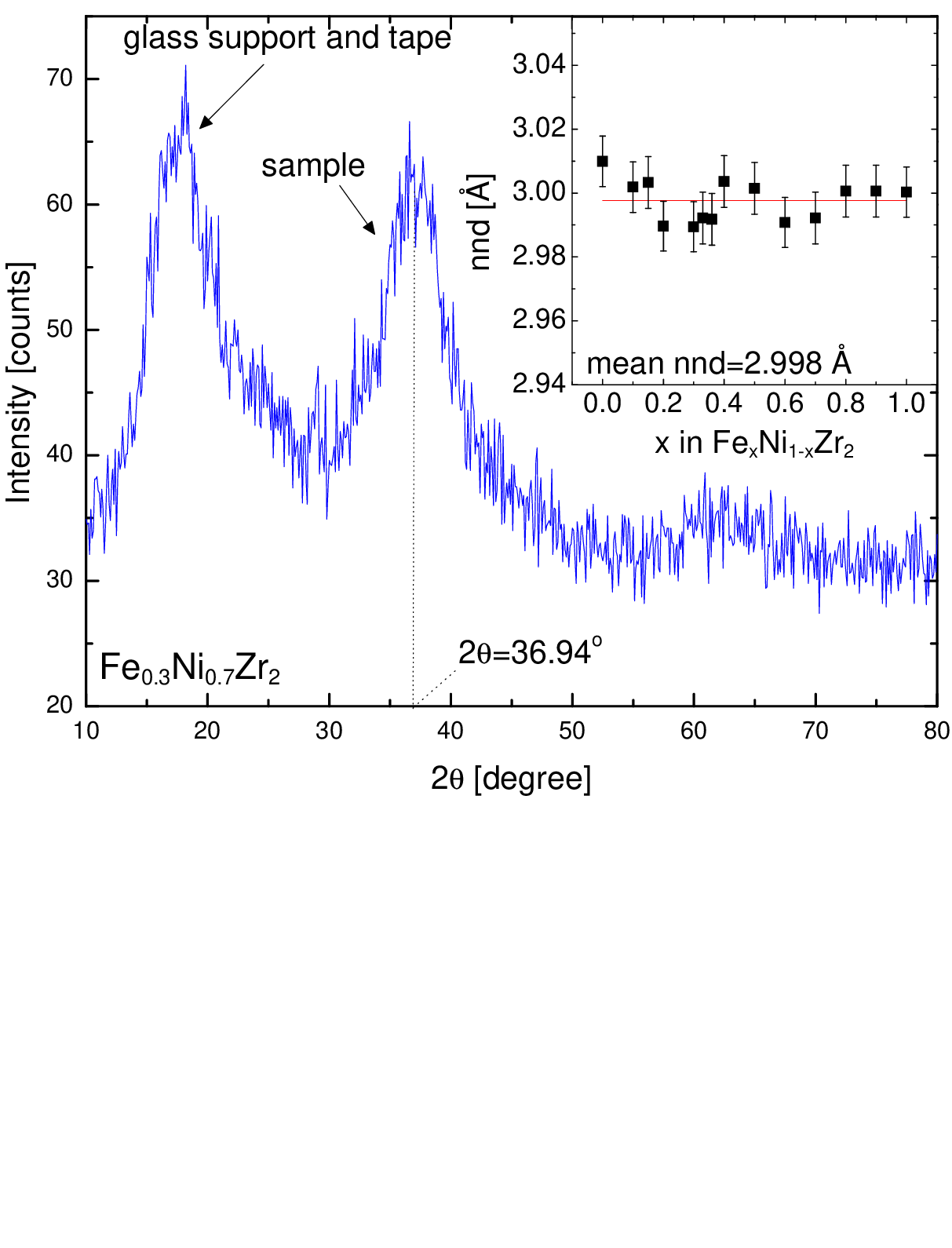}%
\caption{X-ray diffraction spectrum of Fe$_{0.3}$Ni$_{0.7}$Zr$_{2}$ measured
with Cu K$_{\alpha}$ radiation. The first peak is from diffraction from the
glass support. The second peak is from diffraction from the sample. The
position of this peak is determined from a Gaussian fit and the value is used
in Ehrenfest's relation to obtain nnd. Inset: Near-neighbor distance for each
alloy. The red line shows the mean nnd for the whole alloy series.}%
\label{xray}%
\end{center}
\end{figure}

Alloy buttons are prepared by arc-melting appropriate amounts of the elemental
constituents Fe (99.9\%), Ni (99.999\%), Zr (99.95\%) under Ti-gettered argon
atmosphere in order to avoid oxidation. The buttons are re-melted 3 times to
ensure homogeneity. Amorphous ribbons are then prepared by melt-spinning the
alloy buttons. Melt-spinning is performed in 40 kPa helium onto a copper wheel
spinning at 50 m/s which ensures that the rapid cooling rate of 10$^{5}%
$-10$^{6}$ K/s necessary for the formation of the amorphous phase is attained.
The absence of crystallinity was confirmed from the absence of constructive
Bragg peaks in x-ray Cu K$_{\alpha}$ diffraction (Fig.\ \ref{xray}). Indium
contacts are soldered to the samples to permit electrical measurements in the
standard four-probe technique. Resistance measurements are performed with a
resistance bridge providing ac current at 15.9 Hz in a $^{3}$He refrigerator.
The use of a dilution refrigerator was also required for measurements of the
superconducting properties of Fe$_{0.6}$Ni$_{0.4}$Zr$_{2}$ due to the low
T$_{c}$ below 0.3 K. In both the $^{3}$He system and the dilution refrigerator
a superconducting magnet with field capability up to 9 T was used. The temperature is determined from calibrated Cernox and RuO resistors. 

\subsection{Glass structure}

The question of the nature of the ordering in the amorphous phase in the pseudo-binary
Fe$_{x}$Ni$_{1-x}$Zr$_{2}$ series deserves particular attention. In the past,
it has been tacitly assumed that the SRO characterizing
the amorphous structure of these alloys does not change upon substitution of
Ni for Fe because these atoms have very similar sizes. Thereupon, various
studies assuming constant glass structure in these, and similar glasses were
undertaken \cite{Bruning, Mingmao, Yamada, AltounianPRB49, DikeakosJNCS250} to
study the dependence of certain effects on glass composition, independently of
structural change effects. Constant geometrical short range order (GSRO)
across the series of alloys, mainly provided by constant near-neighbor
distances (nnd), is readily verified from the position of the primary
diffraction peak $\theta$ in x-ray diffraction data and from the Ehrenfest
relation \cite{James, Sabet} $\bar{r}=0.6148\lambda/\sin\theta$, the mean
near-neighbor distance $\bar{r}$ is evaluated, as shown in the inset of
Fig.\ \ref{xray}, using $\lambda=1.5405$ \AA , the wavelength of Cu K$_{\alpha}$
radiation. As observed, the nnd in this alloy series is indeed constant, with
a mean nnd of 2.998 \AA , which confirms constant GSRO in these alloys. This
however does not necessarily imply that SRO is constant: since Fe and Ni have different
electronic structures, chemical short range order (CSRO), pertaining mainly to
the atomic species of near-neighbors and their arrangement, cannot be assumed to remain constant.
This question was previously investigated in these alloys using M\"{o}ssbauer
spectroscopy \cite{DikeakosJNCS250} but no change in SRO with x could be
evidenced outside experimental uncertainties. However, some results about superconductivity in this alloy series
point to a transition in SRO \cite{Mfluc}; these will be shown and
discussed later in this article.

\section{Results}

\subsection{Superconducting properties}

\begin{figure}
[ptbh]
\begin{center}
\includegraphics[
trim=0.000000in 0.633982in 0.761774in 0.000000in,
height=4.5403in,
width=3.3399in
]%
{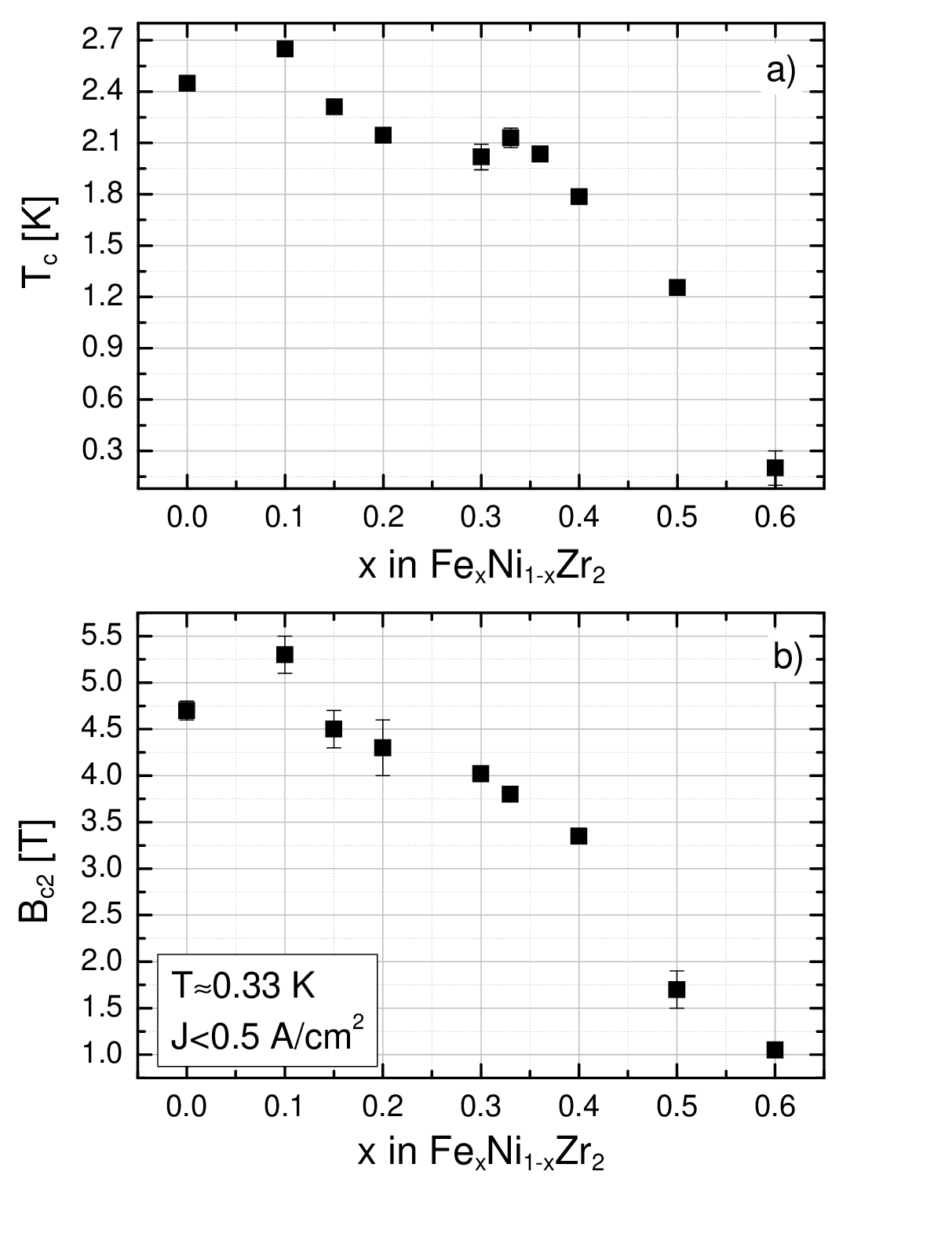}%
\caption{a) Critical temperature and b) Upper critical field as a function of
Fe content x in Fe$_{x}$Ni$_{1-x}$Zr$_{2}$. Error bars are statistical. For
x=0.6, T $<$ 100 mK during measurement of B$_{c2}$.}
\label{TcandBc2}%
\end{center}
\end{figure}

Initial interest in the study of superconductivity in the metallic glasses
Fe$_{x}$Ni$_{1-x}$Zr$_{2}$ was based on the assumed constant glass structure.
Indeed, this would permit a study of the dependence of superconductivity on
alloy composition, and more specifically the influence of spin fluctuations,
independently of structure-dependent effects \cite{AltounianPRB49}. Spin
fluctuations, induced by the presence of magnetic atoms in the alloys, tend to
demote superconductivity by causing spin flips which break Cooper pairs.
Consequently, one expects a suppression of the typical parameters
characterizing superconductivity, such as critical temperature T$_{c}$ and
upper critical field B$_{c2}$ with increasing Fe content. Such a behavior is
witnessed in these alloys, as shown in Fig.\ \ref{TcandBc2}a) and b). As can be
seen, T$_{c}$ of these alloys varies from 2.6 K to about 0.2 K with increasing
Fe content. T$_{c}$ is determined from resistance measurements in zero
magnetic field and defined when the resistance reaches half the normal state
value i.e. at R$_{n}$/2. The values reported for the alloys are from several measurements on up to
five different samples of each alloy composition. Among samples of the same
alloy, typical T$_{c}$ variations observed are smaller than 0.1 K; such a
distribution of T$_{c}$s in an alloy is expected and inherent to the
fabrication process. Indeed, the copper wheel used for melt-spinning becomes
hotter in the process such that not the whole ribbon is cooled at exactly the
same rate and the beginning of the ribbon can show significant differences in
microstructure compared to the end of the ribbon; these differences are then reflected in
superconducting properties. Typically, annealing the samples at a temperature close to the glass transition temperature will remove these differences in microstructure along the length of the ribbon. However, doing this, we nevertheless get a distribution of T$_{c}$s in a single alloy.
 The augmentation of T$_{c}$ from x=0 to 0.1 could
result from an enhancement of the density of states at the Fermi level in
Fe$_{0.1}$Ni$_{0.9}$Zr$_{2}$ compared to NiZr$_{2}$; this idea is supported by
evidence from ultraviolet photoemission spectroscopy on binary alloys of Fe-Zr
and Ni-Zr which have shown that the Fe d band lies closer to the Fermi level
than the Ni d band \cite{OelhafenSSC35}.

Similarly, the upper critical field B$_{c2}$ (Fig.\ \ref{TcandBc2}b)) was
determined from resistance measurements performed as a function of magnetic
field at a fixed temperature between 0.33 and 0.35 K for the $0\leq x\leq0.5$
alloys using a low driving current density $J<0.5$ A/cm$^{2}$. For the $x=0.6$
alloy, the temperature was below 0.1 K. The B$_{c2}$ transition is defined at
R$_{n}$/2 and the values reported are from a few measurements on different
samples. It will be shown later that the B$_{c2}$ transition exhibits large
clockwise hysteresis loops in the $x=0.5$ and $x=0.6$ alloys; in these cases,
the value of B$_{c2}$ reported corresponds to the mean value obtained from up
and down-going field sweeps, i.e. $(B_{c2}^{down}-B_{c2}^{up})/2$. It is found
that B$_{c2}$ decreases from 5.3 T to 1 T with increasing Fe content, and just
like T$_{c}$, B$_{c2}$ increases for the alloy x=0.1 compared to x=0, again
pointing to some promotion of superconductivity by the introduction of a small
amount of Fe.

\begin{table*} \begin{minipage} {\textwidth}

\caption
{Some measured and calculated physical and superconducting parameters.
\label{parameters}}
\begin{ruledtabular}
\begin{tabular}
[c]{||c|c|c|c|c|c|c|c|c||}\hline\hline
Alloy & $\rho_{n}~[\mu\Omega$ m] & $\left.  \frac{dB_{c2}}{dT}\right\vert _{T_{c}}~[T/K]$\footnote{The errors reported on $\left.  \frac{dB_{c2}}{dT}\right\vert
_{T_{c}}$ consider the maximal and minimal slopes that could be obtained
considering systematic errors on T and B$_{c2}$.} & $\lambda\left(  0\right)
~$[$\mu$m]\footnote{Obtained from $\lambda\left(  0\right)  =1.05\times
10^{-3}\left(  \frac{\rho_{n}}{T_{c}}\right)  ^{1/2}$.} & $\xi_{G}\left(
0\right)  $ [nm]\footnote{From $\xi_{G}\left(  0\right)  =1.81\times
10^{-8}\left[  -T_{c}\left\vert \frac{dB_{c2}}{dT}\right\vert _{T_{c}}\right]
^{-1/2}$.} & $\xi_{GL}\left(  0\right)  $ [nm]\footnote{From $\xi_{GL}\left(
0\right)  =\left[  \frac{\Phi_{0}}{2\pi B_{c2}\left(  0\right)  }\right]
^{1/2}$. $B_{c2}\left(  0\right)  $ is obtained from the extrapolation to
$T=0$ of fits to the WHHM theory \cite{WerthamerPR147} of $B_{c2}\left(
T\right)  $ data. } & $\kappa$\footnote{From $\kappa=3.54\times10^{4}\left[  -\rho
_{n}\left\vert \frac{dB_{c2}}{dT}\right\vert _{T_{c}}\right]  ^{1/2}$.} & $l$
[\AA ]\footnote{The \% error on $l$ is computed by considerering the
effect on $l$ of using a free electron-like Fermi surface ratio $S/S_{F}=1$.} & B$_{c1}$ [mT]\footnote{Measured at T $\simeq$ 0.35 K from local magnetization measurements \cite{Mfluc}.}\\\hline\hline
NiZr$_{2}$ & 1.68 $\pm$ 0.02 & -3.0~$\pm$ 0.2 & 0.87 $\pm~5\%$ & 6.7
$\pm~20\%$ & 8.1 $\pm~4\%$ & 96 $\pm~20\%$ & 2.7 $\pm~40\%$ & 0.175 $\pm$ 0.005\\\hline
Fe$_{0.1}$Ni$_{0.9}$Zr$_{2}$ & 1.68 $\pm$ 0.01 & -2.2 $\pm$ 0.4 & 0.84 & 7.5 &
8.0 & 101 & 2.7 & 0.197\\\hline
Fe$_{0.15}$Ni$_{0.85}$Zr$_{2}$ & 1.62 $\pm$ 0.08 & -2.8 $\pm~$0.1 & 0.88 &
7.2 & 8.3 & 106 & 2.9 & N. A.\\\hline
Fe$_{0.2}$Ni$_{0.8}$Zr$_{2}$ & 1.69 $\pm$ 0.01 & -2.4 $\pm$ 0.4 & 0.93 & 8.0 &
8.2 & 99 & 2.8 & 0.295\\\hline
Fe$_{0.3}$Ni$_{0.7}$Zr$_{2}$ & 1.75 $\pm$ 0.02 & -2.8 $\pm$ 0.1 & 0.98 & 7.6 &
8.8 & 103 & 2.7 & 0.210\\\hline
Fe$_{0.33}$Ni$_{0.67}$Zr$_{2}$ & 1.84~$\pm$ 0.08 & -3.2~$\pm$
0.1\footnote{\label{Tsweepnote}Determined from resistance measurements as a
function of magnetic field sweeps at different temperatures.} & 0.98 & 7.0 &
8.7 & 86 & 2.6 & N. A.\\\hline
Fe$_{0.36}$Ni$_{0.64}$Zr$_{2}$ & 1.72 $\pm$ 0.05 & -3.2 $\pm$ 0.1 & 0.97 &
7.1 & N. A. & 83 & 2.7 & N. A.\\\hline
Fe$_{0.4}$Ni$_{0.6}$Zr$_{2}$ & 1.70 $\pm$ 0.02 & -2.6 $\pm$ 0.2 & 1.02 & 8.4 &
9.6 & 131 & 2.8 & 0.279\\\hline
Fe$_{0.5}$Ni$_{0.5}$Zr$_{2}$ (1) & 1.69 $\pm~$0.01 & -2.3~$\pm$
0.1$^{\text{\ref{Tsweepnote}}}$ & 1.22 & 10.1 & 12.3 & 70 & 2.8 & 0.101\\\hline
Fe$_{0.6}$Ni$_{0.4}$Zr$_{2}$ & 1.67 $\pm$ 0.01 & N. A. & 3.04 & N. A. & 16.2 &
N. A. & 2.9 & N. A.\\\hline\hline
\end{tabular}
\end{ruledtabular}
\end{minipage}
\end{table*}

In Table \ref{parameters}, we report various physical and superconducting
parameters of the Fe$_{x}$Ni$_{1-x}$Zr$_{2}$ metallic glasses. The normal
state resistivity $\rho_{n}$ is calculated from resistance measurements
performed at room temperature on long ribbons $(>30~cm)$ such as to minimize
geometry-dependent effects. $\rho_{n}$ of the order 1.68$~\mu\Omega$m is
obtained for all alloys; these values are close to values reported for similar
alloys \cite{AltounianPRB27, KarkutPRB28, Hamed, Ristic}. Since $\rho_{n}$ is
related to the GSRO, the constant $\rho_{n}$ throughout this composition range
brings further confirmation of the constant GSRO. The slope of the upper
critical field as a function of temperature close to T$_{c}$, $\left.
\frac{dB_{c2}}{dT}\right\vert _{T_{c}}$, is determined from a single set of
measurements of the resistance as a function of temperature at different
magnetic fields. We obtain $\left.  \frac{dB_{c2}}{dT}\right\vert _{T_{c}%
}\simeq-2.4~$T/K as typical in amorphous alloys (see for instance
\cite{DombJLTP33, KarkutPRB28, Poon}). We also report values for the
penetration depth $\lambda\left(  0\right)  $, coherence length $\xi
_{G}\left(  0\right)  $ (analogous to BCS $\xi_{0}$ but including correction
due to short mean free path, as indicated by Gor'kov \cite{GorkovSPJETP9,
GorkovSPJETP10b}), and Ginzburg-Landau coherence length $\xi_{GL}\left(
0\right)  $ and GL\ parameter $\kappa$ evaluated from expressions for
superconductors in the dirty limit \cite{KesPRB28}. In NiZr$_{2}$, we obtain $\lambda\left(
0\right)  $= 0.87 $\mu$m and $\xi_{GL}\left(  0\right)  \simeq8.1$ nm; these values generally increase with Fe content to reach 3.04
$\mu$m and 16.2 nm respectively in x=0.6. This means that the vortex size and core increase substantially when going from x=0 to x=0.6. This should have important
effects on vortex pinning properties and vortex-vortex interactions. The GL parameter $\kappa$ is around 80 in
these alloys, thus confirming that they are hard type-II superconductors. The
mean free path is evaluated from
\begin{equation}
l=(3\pi^{2})^{1/3}\left[  e^{2}\rho_{n}\left(  n_{e}^{2/3}\frac{S}{S_{F}%
}\right)  \right]  ^{-1}%
\end{equation}
where $n_{e}$ is the free electron density and $S/S_{F}$ is the ratio of the
area of the free Fermi surface to that of a free electron gas of density
$n_{e}$. Both these quantities are estimated as follows: $n_{e}$ is the ratio
of the average number of electrons per atom outside closed shells to the
atomic volume, i.e. $n_{e}=\left\langle \frac{e}{a}\right\rangle V_{0}^{-1}$.
In this manner we have $2.9\times10^{29}$ m$^{-3}\leq n_{e}\leq3.2\times
10^{29}$ m$^{-3}$, with $n_{e}$ decreasing with increasing Fe content. We also
use $S/S_{F}=0.6$ as in Ref.\cite{KarkutPRB28}; although if we were to use
$S/S_{F}=1$ as for a free electron Fermi surface the value of $l$ would not
change by an order of magnitude. As a result, we obtain $l\simeq2.8~$\AA which is very close to the mean interatomic distance (nnd $=$ 2.998 \AA ) in
these amorphous alloys. As a result of the short mean free path, we obtain a
dirtiness parameter\footnote{The dirtiness parameter is usually computed from
the ratio of the BCS coherence length $\xi_{0}$ to the mean free path.
However, we use the experimentally determined $\xi_{G}\left(  0\right)  $
because it represents the real coherence length of our samples with
consideration for the short mean free path in the dirty limit.} $\xi
_{G}\left(  0\right)  /l$ above 20 which effectively confirms that these
amorphous alloys are in the dirty limit. We also evaluate the electron-phonon
coupling parameter $\lambda_{ep}\simeq0.6$ using the McMillan equation
\cite{McMillanPR167}%
\begin{equation}
\lambda_{ep}=\frac{1.04+\mu^{\ast}\ln\left(  \Theta_{D}/1.45T_{c}\right)
}{\left(  1-0.62\mu^{\ast}\right)  \ln\left(  \Theta_{D}/1.45T_{c}\right)
-1.04}%
\end{equation}
where we have used the Coulomb interaction parameter $\mu^{\ast}=0.13$ for
polyvalent transition metals \cite{McMillanPR167} and the Debye temperature
$\Theta_{D}=192.5$ K as evaluated for NiZr$_{2}$ according to
Ref.\cite{AltounianPRB27}. This $\lambda_{ep}$ makes the Fe$_{x}$Ni$_{1-x}
$Zr$_{2}$ metallic glasses in the weak to intermediate coupling regime.

\subsection{Structure inhomogeneity and CSRO}

Several physical properties of amorphous alloys depend on CSRO, for instance:
the Curie temperature, the temperature coefficient of resistivity and
superconducting properties. In the case of superconductivity, this is because
structural order directly influences the electron-phonon coupling parameter, resulting in a
modification of T$_{c}$ for instance. Structural order can also influence
superconductivity by acting on vortex-pin interactions, in
particular defects, impurities and inhomogeneities, and can thus define the
current-carrying capacities of the superconductor. For instance, it is the
absence of long-range order in amorphous alloys which mainly determine their
weak vortex pinning properties. Effects of structural inhomogeneity
are also commonly observed in superconducting properties, such as wide
B$_{c2}$ or T$_{c}$ transitions. In this section, we discuss some evidences of
structural inhomogeneity in the Fe$_{x}$Ni$_{1-x}$Zr$_{2}$ with a large Fe
content 0.4 $\leq x\leq$ 0.6 obtained from resistance and magnetization
measurements in the superconducting state.

\subsubsection{Width of transition}

\begin{figure}
[ptbh]
\begin{center}
\includegraphics[
trim=0.000000in 6.094917in 0.000000in 0.000000in,
height=1.8092in,
width=3.2214in
]%
{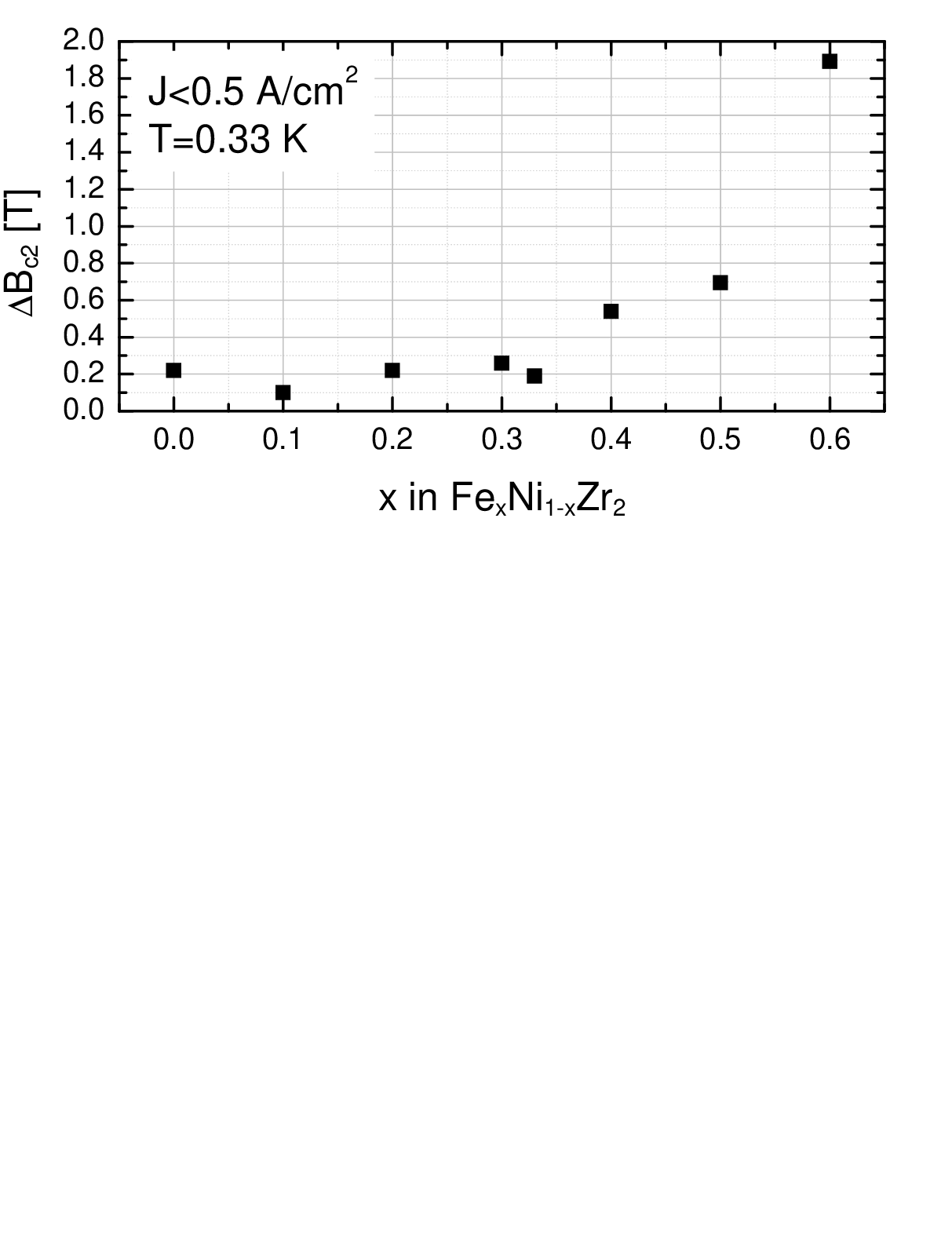}%
\caption{Width of the B$_{c2}$ transition for different Fe$_{x}$Ni$_{1-x}%
$Zr$_{2}$ alloys. In x=0.6, this includes both transitions 1 and 2.}%
\label{Bc2width}%
\end{center}
\end{figure}

A first obvious sign of the growth of structural inhomogeneity with Fe content
in these alloys is provided by an increase in the width of the B$_{c2}$
transition $\Delta$B$_{c2}\equiv \ $B(0.9R$_{n}$)-B(0.1R$_{n}$), as shown in
Fig.\ \ref{Bc2width}. A double-step transition is even observed in some of the
alloys x=0.5 and 0.6 (Fig.\ \ref{RvsBall}), an indisputable indication of inhomogeneity.%

\subsubsection{Clockwise hysteresis}
\begin{figure}
[ptbh]
\begin{center}
\includegraphics[
trim=0.000000in 5.714316in 0.000000in 0.000000in,
height=2.0505in,
width=3.3702in
]%
{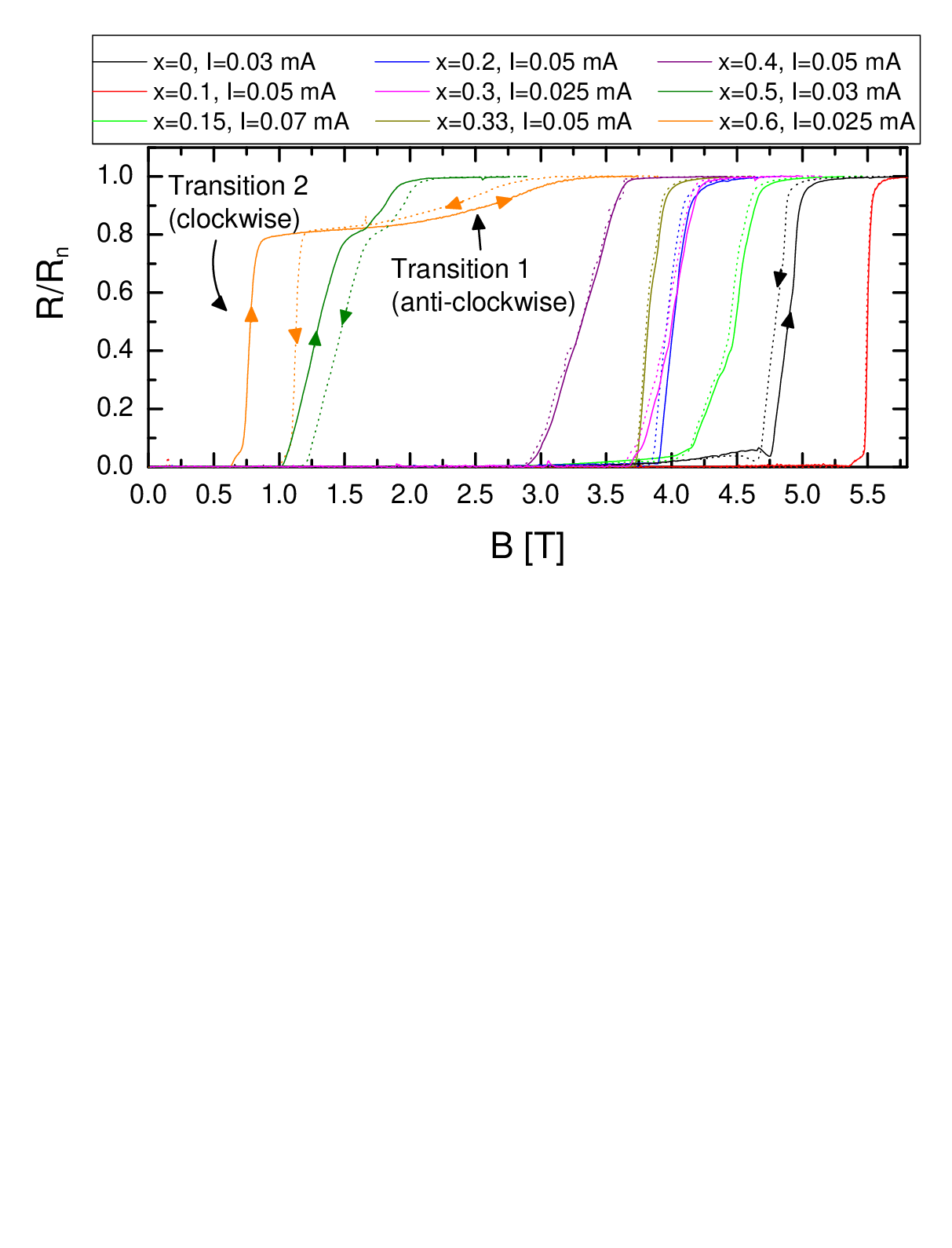}%
\caption{Resistance as a function of magnetic field for different alloys. The
solid and dotted lines are for increasing and decreasing B sweep respectively.
The magnetic field was swept at a rate of 0.0147 T/s and the temperature was
below 0.35 K.}%
\label{RvsBall}%
\end{center}
\end{figure}

In type-II superconductors, the B$_{c2}$ transition between the normal state
and the superconducting state often exhibits hysteresis due to Joule heating where more power is dissipated in the normal state than in the superconducting state, and to flux pinning
and trapping. The hysteresis loop is then counterclockwise, i.e. the
B$_{c2}$ transition is higher upon increasing the magnetic field than when
decreasing it. These types of hysteresis loops are observed here for alloys
with $0\leq x\leq0.4$ (Fig.\ \ref{RvsBall}) in which the B$_{c2}$ transition is
slightly lower upon decreasing (dotted lines) the magnetic field than upon
increasing (solid lines) it. This can be contrasted to the large clockwise hysteresis loops seen at B$_{c2}$ in the alloys $x=0.5$ and $x=0.6$. All the $x=0.5$
samples measured (7) show this wide clockwise hysteresis loop, although only 2
of them show the double-step transition. The only $x=0.6$ sample measured
shows a very broad B$_{c2}$ transition about 2 T wide including the two steps.
A reversal of the hysteresis loop direction in x $=$ 0.6 is observed between transition 1
and 2 as identified in the figure; the uppermost transition exhibits the usual
counterclockwise hysteresis loops.

\begin{figure}
[ptbh]
\begin{center}
\includegraphics[
trim=0.000000in 4.568273in 0.000000in 0.000000in,
height=2.4128in,
width=3.2214in
]%
{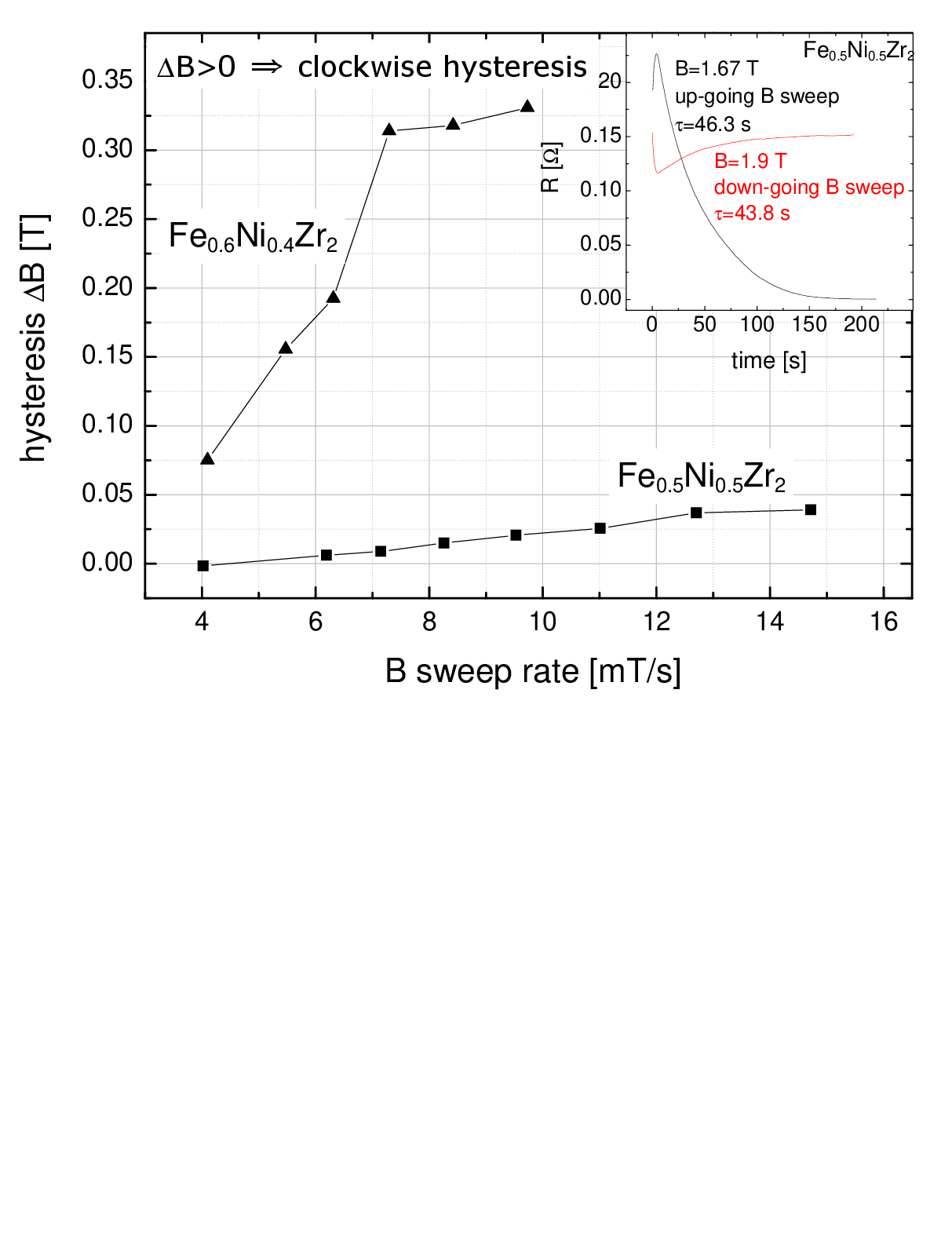}%
\caption{Width of clockwise hysteresis loop as a function of magnetic field
sweep rate. $\Delta$B is defined as the difference in magnetic field when the
resistance reaches 0.5R$_{n}$ for down-going (B$_{c2}^{down}$) and up-going
(B$_{c2}^{up}$) field sweep, i.e. $\Delta$B$=\left(  \text{B}_{c2}%
^{down}-\text{B}_{c2}^{up}\right)  /2$. Only the lowermost transition
(transition 2) is considered here in x=0.6. For Fe$_{0.5}$Ni$_{0.5}$Zr$_{2}$,
T=0.34 K and I=30 $\mu$A. For Fe$_{0.6}$Ni$_{0.4}$Zr$_{2}$, T$<$0.1 K and I=5
$\mu$A. The line is a guide for the eye. Inset: Resistance as a function of
time when B field sweep is paused in the middle of the B$_{c2}$ transition.
I=0.03 mA, T=0.33 K.}%
\label{sweeprate}%
\end{center}
\end{figure}

We have resistively measured the B$_{c2}$ transition in x=0.5 and 0.6 for
different B field sweep rates. The results, presented in Fig.\ \ref{sweeprate},
show an increase in the size of the hysteresis loops with increasing B sweep
rate. This dependence on sweep rate provides evidence that a dynamical
process, such as vortex motion, is at the origin of the hysteresis loops. This
is further supported by the data shown in the inset of Fig.\ \ref{sweeprate},
which shows the resistance as a function of time when the magnetic field sweep
is paused in the middle of the B$_{c2}$ transition during an increasing and a
decreasing magnetic field sweep. Time $t=0$ corresponds to the moment when the
field sweep is paused. As can be seen, after $t=0$ the resistance initially
keeps increasing (decreasing) over a short period of time for increasing
(decreasing) B sweep, but it eventually reverses and decreases (increases)
back to zero (a value close to what it was before the sweep was paused).
Fitting these time dependences to an exponential obtains a time constant
$\tau$ of 46.3 s and 43.8 s for the up and down-going field sweeps
respectively, indicating a slow dynamical process. On the contrary to what is
observed in magnetic field sweeps, no clockwise hysteresis loops are observed
in temperature sweeps across T$_{c}$ in a fixed magnetic field (data not
shown). This brings further confirmation that the clockwise hysteresis loops
are governed by a dynamical process involving vortex motion rather than some
phase transition.

In the literature, simulations of V-I characteristics in superconductors
with inhomogeneous pinning potentials show clockwise hysteresis loops
\cite{LiuPRB66, Xu}. Such hysteresis loops result from dynamical effects and
interplay between vortex trapping in the strong and weak pinning regions. As
such, the size (or width) of the hysteresis loops is seen to depend on the
sweep rate of the external variable with respect to which the loop is
observed; in Refs. \cite{LiuPRB66} and \cite{Xu}, this is the driving current and
driving force. As is the case in our B field-induced hysteresis loops, the
size of the loops increases in these Refs. \cite{LiuPRB66, Xu} for faster driving force
sweep speeds.

\begin{figure}
[ptbh]
\begin{center}
\includegraphics[
height=2.1491in,
width=3.2396in
]%
{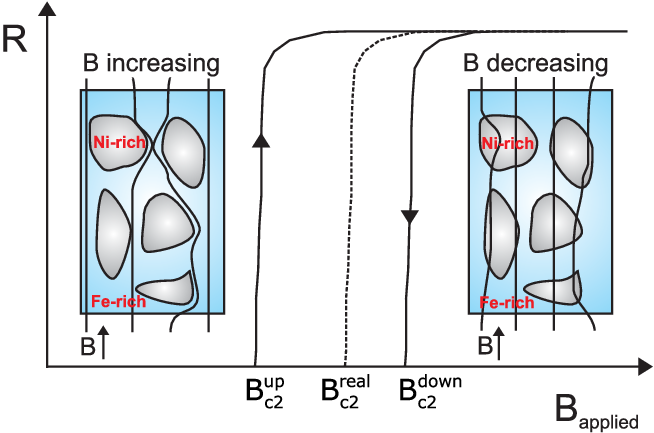}%
\caption{Schematic representation of the resistive B$_{c2}$ transition for up
and down-going field sweeps with corresponding inhomogeneous vortex
distribution in the superconductor. For increasing B field, more vortices pass
through the weakly-pinned Fe-rich phase, avoiding the Ni-rich regions and the
B$_{c2}^{up}$ transition appears lower than it should be if the fluxes were
homogeneously distributed in the whole sample (B$_{c2}^{\text{real}}$). The
inverse phenomena occurs upon decreasing B field, with vortices tending to
remain in the more strongly pinned Ni-rich clusters such that B$_{c2}^{down}$
appears higher than B$_{c2}^{\text{real}}$.}%
\label{clusterdrawing}%
\end{center}
\end{figure}

Based on these simulation results and our data, we propose that the anomalous
clockwise hysteresis loops observed at the B$_{c2}$ transition here arise due
to the presence of inhomogeneities, some having stronger and weaker pinning
properties, and thus resulting in an inhomogeneous distribution of vortices in
the superconductors. If we picture the sample as being composed of regions
having stronger pinning properties than the surrounding medium, and if this
medium also provides the connected path across the sample, we can explain the
appearance of clockwise hysteresis loops in magnetic field sweeps and their
absence in temperature sweeps as follows: As B is increased from 0, vortices
will first penetrate in the main connected phase since it has the lowest
pinning properties and thus a lower energy barrier against flux entry. Then,
due to the elasticity of the vortex lattice, as the magnetic field is
increased further it will be energetically more favorable for the vortices to
bend around the strong pinning regions \cite{JiPRB47} and remain in the main
connected phase. This will result in an inhomogeneous distribution of magnetic
flux in the superconductor with a larger flux density being present in the
main connected phase, i.e. the phase of which the superconducting properties
are measured in resistance measurements. As a consequence, the B$_{c2}$
transition appears lower upon increasing the magnetic field than it would be
if the fluxes were homogeneously distributed throughout the sample. A
schematic representation of this process is presented in
Fig.\ \ref{clusterdrawing}. On the contrary, upon decreasing the magnetic field
from above B$_{c2}$, fluxes tend to stay trapped in the strong-pinning regions
but easily leave the weak-pinning main phase which again results in an
inhomogeneous distribution of vortices in the sample, but this time with the
lowest vortex density in the main phase. This results in a
resistively-measured B$_{c2}$ transition upon decreasing the magnetic field
that is higher than it should be. These two processes then result in clockwise
hysteresis loops. Similar phenomena have been observed previously in granular
superconductors \cite{JiPRB47, Kilic, dosSantosPhysicaC391} and inhomogeneous
superconductors \cite{PoonPRB25, LiuPRB66, Xu}.

This model explains why we witness an increase in the width of hysteresis
loops with increasing B sweep rate: for slower sweep rate the vortices have
more time to penetrate into the strong-pinning grains or to diffuse into the
weak-pinning regions upon increasing and decreasing the magnetic field before
the measurement is taken. Therefore, after we apply a certain field, and by
the time we take the measurement, the vortex distribution has reached a more
homogeneous configuration for slower sweep speeds and results in smaller
hysteresis loops. A similar observation was made by Liu \textit{et al.}
\cite{LiuPRB66} and Xu \textit{et al.} \cite{Xu} from numerical simulations
and experimental measurements of flux creep in a superconductor with
inhomogeneous pinning properties. In these cases however, the size of
hysteresis loops was observed to increase with increasing driving current and
driving force sweep rate, but the result is equivalent: changing the sweep
rate amounts to changing the observation time window. Namely, for a slow sweep
rate, our observation window is too late to observe the large inhomogeneity in
the flux distribution.

The results presented in the inset of Fig.\ \ref{sweeprate} also lead to the
conclusion that the vortex distribution is inhomogeneous in these alloys. For
instance, when the field is paused in an increasing magnetic field sweep, the
resistance initially keeps increasing because fluxes easily and rapidly enter
the weak-pinning phase; however it eventually starts decreasing as no more
fluxes are added (paused B field) and the fluxes in the weakly-pinned regions
start to diffuse in the strong pinning grains, thus yielding a more
homogeneous vortex distribution which brings the observed B$_{c2}$ transition
closer to the real value. The opposite takes place when the field is paused in
a decreasing B sweep; the resistance change over approximately the same period
of time is however smaller and indicates that the flux distribution is more
homogeneous at B$_{c2}$ in a decreasing B sweep than in an increasing one.

According to this model, no clockwise hysteresis loops are expected in
temperature sweeps performed in a fixed external field, because in this case,
the vortex density is fixed and its distribution across the sample remains the
same as the temperature is swept up and down.

\subsubsection{Magnetization fluctuations}

\begin{figure}
[ptbh]
\begin{center}
\includegraphics[
trim=0.000000in 8.132564in 3.045481in 0.000000in,
height=1.6535in,
width=3.339in
]%
{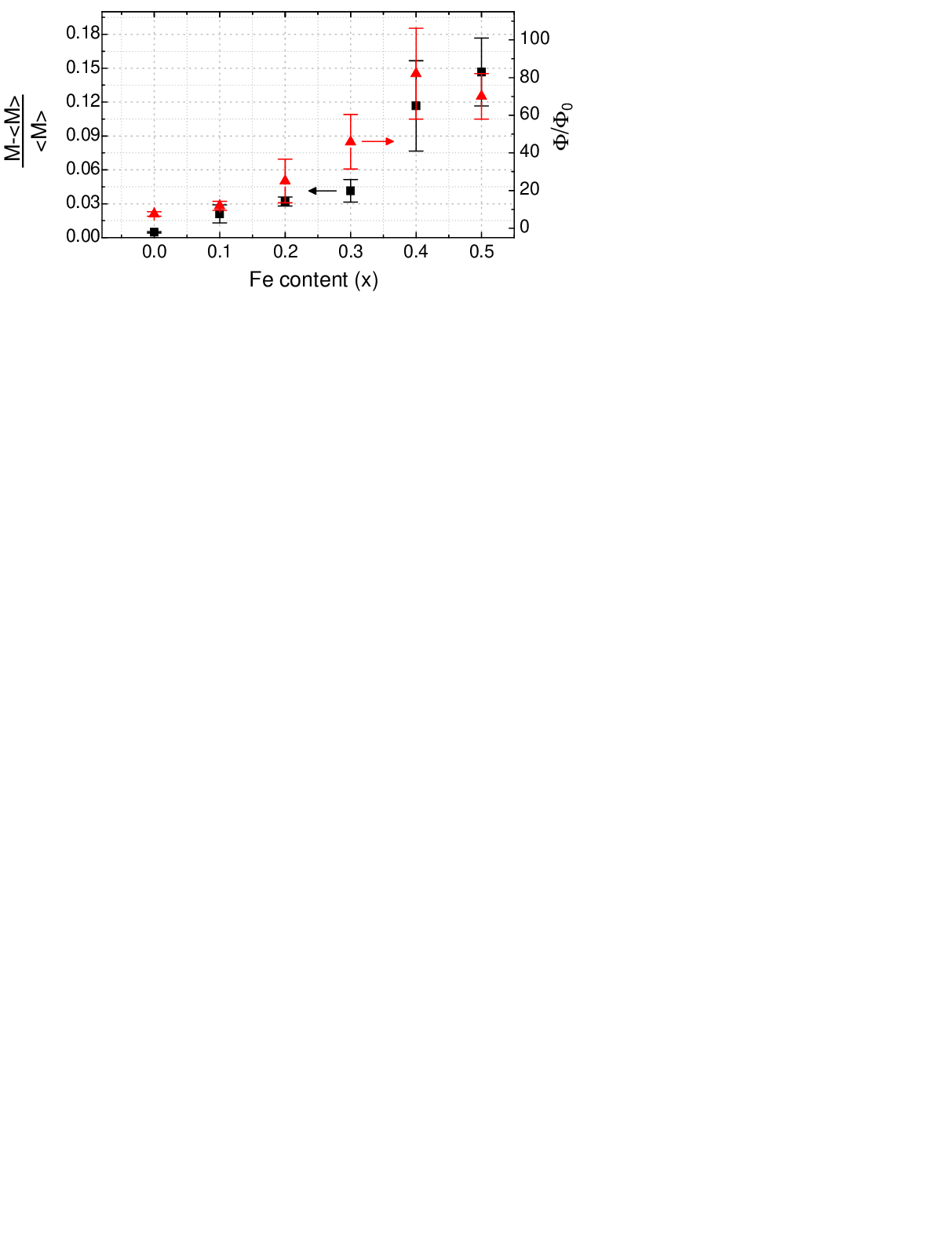}%
\caption{Black squares: Relative size of magnetization fluctuations for
different glasses Fe$_{x}$Ni$_{1-x}$Zr$_{2}$. Red triangles: Number of
vortices in clusters related to the magnetization fluctuations.}%
\label{Sizefluc}%
\end{center}
\end{figure}

As described in detail elsewhere \cite{myPhDthesis, Mfluc}, local
magnetization measurements were performed on these metallic glasses (0 $\leq
x\leq$ 0.5) using a 2-dimensional electron gas (2DEG) Hall probe. Fluctuations
in the magnetization, increasing with Fe content
(Fig.\ \ref{Sizefluc}), were observed and analyzed to reveal the presence of
large vortex clusters of over 70 vortices in the superconductors with a large
Fe content $x > 0.4$. It was also argued that the vortex bundles likely arise in these alloys
because they are composed of two phases having different SRO, and thus
different pinning properties. In all alloys with $x>0$, both a Fe-rich and an
Ni-rich phase exist, as evidenced from larger magnetization fluctuations in
$x=0.1$ compared to $x=0$, but the regions of the Fe-rich phase become larger
and more numerous in $x>0.4$; these results point to the existence of a
structural phase transition close to $x=0.4$, with alloys $x<0.4$ having
mostly NiZr$_{2}$-like SRO and alloys with $x>0.4$ having mostly FeZr$_{2}%
$-like SRO.

Summarizing the results presented in this section, many evidences indicate
phase separation into Fe-rich and Ni-rich regions in the Fe$_{x}$Ni$_{1-x}%
$Zr$_{2}$ metallic glasses, particularly in the compositions with a large Fe
content. Both the gradual increase in the width of the B$_{c2}$ transition
(Fig.\ \ref{Bc2width}) and increasing size of magnetization fluctuations
(Fig.\ \ref{Sizefluc}) with x point to an augmentation of the presence of
inhomogeneities, most likely Fe-rich regions. The very broad $\Delta$B$_{c2}$,
double-step transition, and appearance of large clockwise hysteresis loops in
$x=0.5$ and $0.6$ further supports this idea and provides indication that
Fe-rich regions have become critically large in these alloys, but some Ni-rich regions remain. 

\section{Conclusions}

In summary, we have presented the Fe content dependence of superconductivity in the metallic glasses Fe$_{x}$Ni$_{1-x}$Zr$_{2}$. As
expected due to the augmentation of spin fluctuations, T$_{c}$ and B$_{c2}$
decrease with increasing x; this decrease becomes very pronounced in x=0.6. Important progress was also made in our understanding of short range order in this amorphous alloy series. While we have shown that GSRO remains constant across the series as expected, numerous evidence, such as
an increase in B$_{c2}$ transition width with x, the appearance of double-step
transitions and large clockwise hysteresis loops, and the observation of large
fluctuations in magnetization in high Fe containing alloys, point to the
conclusion that two phases having different CSRO exist in these alloys . According to these results,
a Fe-rich phase having SRO resembling that of FeZr$_{2}$ appears with a Fe
content as low as x=0.1, but becomes critical around x=0.5 where important
effects on superconductivity are witnessed. These results emphasize the
extreme sensitivity of superconductivity, and more specifically of vortices,
as a probe of the microstructure of materials.


\begin{thebibliography}{35}
\expandafter\ifx\csname natexlab\endcsname\relax\def\natexlab#1{#1}\fi
\expandafter\ifx\csname bibnamefont\endcsname\relax
  \def\bibnamefont#1{#1}\fi
\expandafter\ifx\csname bibfnamefont\endcsname\relax
  \def\bibfnamefont#1{#1}\fi
\expandafter\ifx\csname citenamefont\endcsname\relax
  \def\citenamefont#1{#1}\fi
\expandafter\ifx\csname url\endcsname\relax
  \def\url#1{\texttt{#1}}\fi
\expandafter\ifx\csname urlprefix\endcsname\relax\def\urlprefix{URL }\fi
\providecommand{\bibinfo}[2]{#2}
\providecommand{\eprint}[2][]{\url{#2}}

\bibitem[{\citenamefont{B\"{u}ckel and Hilsch}(1954)}]{BuckelZphys138}
\bibinfo{author}{\bibfnamefont{W.}~\bibnamefont{B\"{u}ckel}} \bibnamefont{and}
  \bibinfo{author}{\bibfnamefont{R.}~\bibnamefont{Hilsch}},
  \bibinfo{journal}{Z. Phys.} \textbf{\bibinfo{volume}{138}},
  \bibinfo{pages}{109} (\bibinfo{year}{1954}).

\bibitem[{\citenamefont{Collver and Hammond}(1973)}]{CollverPRL30}
\bibinfo{author}{\bibfnamefont{M.~M.} \bibnamefont{Collver}} \bibnamefont{and}
  \bibinfo{author}{\bibfnamefont{R.~H.} \bibnamefont{Hammond}},
  \bibinfo{journal}{Phys. Rev. Lett.} \textbf{\bibinfo{volume}{30}},
  \bibinfo{pages}{92} (\bibinfo{year}{1973}).

\bibitem[{\citenamefont{Altounian and Strom-Olson}(1983)}]{AltounianPRB27}
\bibinfo{author}{\bibfnamefont{Z.}~\bibnamefont{Altounian}} \bibnamefont{and}
  \bibinfo{author}{\bibfnamefont{J.~O.} \bibnamefont{Strom-Olson}},
  \bibinfo{journal}{Phys. Rev. B} \textbf{\bibinfo{volume}{27}},
  \bibinfo{pages}{4149} (\bibinfo{year}{1983}).

\bibitem[{\citenamefont{Sabouri-Ghomi and Altounian}(1996)}]{Sabouri}
\bibinfo{author}{\bibfnamefont{M.}~\bibnamefont{Sabouri-Ghomi}}
  \bibnamefont{and}
  \bibinfo{author}{\bibfnamefont{Z.}~\bibnamefont{Altounian}},
  \bibinfo{journal}{J. Non-Crys. Sol.} \textbf{\bibinfo{volume}{205-207}},
  \bibinfo{pages}{692} (\bibinfo{year}{1996}).

\bibitem[{\citenamefont{Trudeau and Cochrane}(1990)}]{Trudeau}
\bibinfo{author}{\bibfnamefont{M.~L.} \bibnamefont{Trudeau}} \bibnamefont{and}
  \bibinfo{author}{\bibfnamefont{R.~W.} \bibnamefont{Cochrane}},
  \bibinfo{journal}{Phys. Rev. B} \textbf{\bibinfo{volume}{41}},
  \bibinfo{pages}{10535} (\bibinfo{year}{1990}).

\bibitem[{\citenamefont{Risti\'{c} et~al.}(1997)\citenamefont{Risti\'{c},
  Marohni\'{c}, and Babi\'{c}}}]{Ristic}
\bibinfo{author}{\bibfnamefont{R.}~\bibnamefont{Risti\'{c}}},
  \bibinfo{author}{\bibfnamefont{v.}~\bibnamefont{Marohni\'{c}}},
  \bibnamefont{and}
  \bibinfo{author}{\bibfnamefont{E.}~\bibnamefont{Babi\'{c}}},
  \bibinfo{journal}{Mater. Sci. Eng.} \textbf{\bibinfo{volume}{A226-228}},
  \bibinfo{pages}{1060} (\bibinfo{year}{1997}).

\bibitem[{\citenamefont{Hamed et~al.}(1995)\citenamefont{Hamed, Razavi, Bose,
  and Startseva}}]{Hamed}
\bibinfo{author}{\bibfnamefont{F.}~\bibnamefont{Hamed}},
  \bibinfo{author}{\bibfnamefont{F.~S.} \bibnamefont{Razavi}},
  \bibinfo{author}{\bibfnamefont{S.~K.} \bibnamefont{Bose}}, \bibnamefont{and}
  \bibinfo{author}{\bibfnamefont{T.}~\bibnamefont{Startseva}},
  \bibinfo{journal}{Phys. Rev. B} \textbf{\bibinfo{volume}{52}},
  \bibinfo{pages}{9674} (\bibinfo{year}{1995}).

\bibitem[{\citenamefont{Flodin et~al.}(1986)\citenamefont{Flodin, Hedman, and
  Rapp}}]{Flodin}
\bibinfo{author}{\bibfnamefont{M.}~\bibnamefont{Flodin}},
  \bibinfo{author}{\bibfnamefont{L.}~\bibnamefont{Hedman}}, \bibnamefont{and}
  \bibinfo{author}{\bibfnamefont{O.}~\bibnamefont{Rapp}},
  \bibinfo{journal}{Phys. Rev. B} \textbf{\bibinfo{volume}{34}},
  \bibinfo{pages}{4558} (\bibinfo{year}{1986}).

\bibitem[{\citenamefont{Karkut and Hake}(1983)}]{KarkutPRB28}
\bibinfo{author}{\bibfnamefont{M.~G.} \bibnamefont{Karkut}} \bibnamefont{and}
  \bibinfo{author}{\bibfnamefont{R.~R.} \bibnamefont{Hake}},
  \bibinfo{journal}{Phys. Rev. B} \textbf{\bibinfo{volume}{28}},
  \bibinfo{pages}{1396} (\bibinfo{year}{1983}).

\bibitem[{\citenamefont{Altounian et~al.}(1994)\citenamefont{Altounian, Dantu,
  and Dikeakos}}]{AltounianPRB49}
\bibinfo{author}{\bibfnamefont{Z.}~\bibnamefont{Altounian}},
  \bibinfo{author}{\bibfnamefont{S.~V.} \bibnamefont{Dantu}}, \bibnamefont{and}
  \bibinfo{author}{\bibfnamefont{M.}~\bibnamefont{Dikeakos}},
  \bibinfo{journal}{Phys. Rev. B} \textbf{\bibinfo{volume}{49}},
  \bibinfo{pages}{8621} (\bibinfo{year}{1994}).

\bibitem[{\citenamefont{Dikeakos et~al.}(1999)\citenamefont{Dikeakos,
  Altounian, Ryan, and Kwon}}]{DikeakosJNCS250}
\bibinfo{author}{\bibfnamefont{M.}~\bibnamefont{Dikeakos}},
  \bibinfo{author}{\bibfnamefont{Z.}~\bibnamefont{Altounian}},
  \bibinfo{author}{\bibfnamefont{D.~H.} \bibnamefont{Ryan}}, \bibnamefont{and}
  \bibinfo{author}{\bibfnamefont{S.~J.} \bibnamefont{Kwon}},
  \bibinfo{journal}{J. Non-Cryst. Solids} \textbf{\bibinfo{volume}{250-252}},
  \bibinfo{pages}{637} (\bibinfo{year}{1999}).

\bibitem[{\citenamefont{Hilke et~al.}(2003)\citenamefont{Hilke, Reid, Gagnon,
  and Altounian}}]{HilkePRL91}
\bibinfo{author}{\bibfnamefont{M.}~\bibnamefont{Hilke}},
  \bibinfo{author}{\bibfnamefont{S.}~\bibnamefont{Reid}},
  \bibinfo{author}{\bibfnamefont{R.}~\bibnamefont{Gagnon}}, \bibnamefont{and}
  \bibinfo{author}{\bibfnamefont{Z.}~\bibnamefont{Altounian}},
  \bibinfo{journal}{Phys. Rev. Lett.} \textbf{\bibinfo{volume}{91}},
  \bibinfo{pages}{127004} (\bibinfo{year}{2003}).

\bibitem[{\citenamefont{Lefebvre et~al.}(2006)\citenamefont{Lefebvre, Hilke,
  Gagnon, and Altounian}}]{LefebvrePRB74}
\bibinfo{author}{\bibfnamefont{J.}~\bibnamefont{Lefebvre}},
  \bibinfo{author}{\bibfnamefont{M.}~\bibnamefont{Hilke}},
  \bibinfo{author}{\bibfnamefont{R.}~\bibnamefont{Gagnon}}, \bibnamefont{and}
  \bibinfo{author}{\bibfnamefont{Z.}~\bibnamefont{Altounian}},
  \bibinfo{journal}{Phys. Rev. B} \textbf{\bibinfo{volume}{74}},
  \bibinfo{eid}{174509} (\bibinfo{year}{2006}).

\bibitem[{\citenamefont{Lefebvre
  et~al.}(2008{\natexlab{a}})\citenamefont{Lefebvre, Hilke, and
  Altounian}}]{LefebvrePRB78}
\bibinfo{author}{\bibfnamefont{J.}~\bibnamefont{Lefebvre}},
  \bibinfo{author}{\bibfnamefont{M.}~\bibnamefont{Hilke}}, \bibnamefont{and}
  \bibinfo{author}{\bibfnamefont{Z.}~\bibnamefont{Altounian}},
  \bibinfo{journal}{Phys. Rev. B} \textbf{\bibinfo{volume}{78}},
  \bibinfo{eid}{134506} (\bibinfo{year}{2008}{\natexlab{a}}).

\bibitem[{\citenamefont{Br\"{u}ning et~al.}(1987)\citenamefont{Br\"{u}ning,
  Altounian, and Str\"{o}m-Olsen}}]{Bruning}
\bibinfo{author}{\bibfnamefont{R.}~\bibnamefont{Br\"{u}ning}},
  \bibinfo{author}{\bibfnamefont{Z.}~\bibnamefont{Altounian}},
  \bibnamefont{and} \bibinfo{author}{\bibfnamefont{J.~O.}
  \bibnamefont{Str\"{o}m-Olsen}}, \bibinfo{journal}{Journal of Applied Physics}
  \textbf{\bibinfo{volume}{62}}, \bibinfo{pages}{3633} (\bibinfo{year}{1987}).

\bibitem[{\citenamefont{Mao and Altounian}(1987)}]{Mingmao}
\bibinfo{author}{\bibfnamefont{M.}~\bibnamefont{Mao}} \bibnamefont{and}
  \bibinfo{author}{\bibfnamefont{Z.}~\bibnamefont{Altounian}},
  \bibinfo{journal}{J. Non-Cryst. Solids} \textbf{\bibinfo{volume}{205-207}},
  \bibinfo{pages}{633} (\bibinfo{year}{1987}).

\bibitem[{\citenamefont{Yamada et~al.}(1988)\citenamefont{Yamada, Itoh, and
  Mizutani}}]{Yamada}
\bibinfo{author}{\bibfnamefont{Y.}~\bibnamefont{Yamada}},
  \bibinfo{author}{\bibfnamefont{Y.}~\bibnamefont{Itoh}}, \bibnamefont{and}
  \bibinfo{author}{\bibfnamefont{U.}~\bibnamefont{Mizutani}},
  \bibinfo{journal}{Mater. Sci. Eng.} \textbf{\bibinfo{volume}{99}},
  \bibinfo{pages}{289} (\bibinfo{year}{1988}).

\bibitem[{\citenamefont{James}(1962)}]{James}
\bibinfo{author}{\bibfnamefont{R.}~\bibnamefont{James}},
  \emph{\bibinfo{title}{Optical Principles of the Diffraction of X-rays}}
  (\bibinfo{publisher}{Cornell University Press}, \bibinfo{year}{1962}).

\bibitem[{\citenamefont{Sabet-Sharghi et~al.}(1994)\citenamefont{Sabet-Sharghi,
  Altounian, and Muir}}]{Sabet}
\bibinfo{author}{\bibfnamefont{R.}~\bibnamefont{Sabet-Sharghi}},
  \bibinfo{author}{\bibfnamefont{Z.}~\bibnamefont{Altounian}},
  \bibnamefont{and} \bibinfo{author}{\bibfnamefont{W.~B.} \bibnamefont{Muir}},
  \bibinfo{journal}{Journal of Applied Physics} \textbf{\bibinfo{volume}{75}},
  \bibinfo{pages}{4438} (\bibinfo{year}{1994}).

\bibitem[{\citenamefont{Lefebvre
  et~al.}(2008{\natexlab{b}})\citenamefont{Lefebvre, Hilke, Altounian, West,
  and Pfeiffer}}]{Mfluc}
\bibinfo{author}{\bibfnamefont{J.}~\bibnamefont{Lefebvre}},
  \bibinfo{author}{\bibfnamefont{M.}~\bibnamefont{Hilke}},
  \bibinfo{author}{\bibfnamefont{Z.}~\bibnamefont{Altounian}},
  \bibinfo{author}{\bibfnamefont{K.~W.} \bibnamefont{West}}, \bibnamefont{and}
  \bibinfo{author}{\bibfnamefont{L.~N.} \bibnamefont{Pfeiffer}}
  (\bibinfo{year}{2008}{\natexlab{b}}), \bibinfo{note}{submitted for
  publication in Phys. Rev. B}.

\bibitem[{\citenamefont{Oelhafen et~al.}(1980)\citenamefont{Oelhafen, Hauser,
  and G\"{u}ntherodt}}]{OelhafenSSC35}
\bibinfo{author}{\bibfnamefont{P.}~\bibnamefont{Oelhafen}},
  \bibinfo{author}{\bibfnamefont{E.}~\bibnamefont{Hauser}}, \bibnamefont{and}
  \bibinfo{author}{\bibfnamefont{H.-J.} \bibnamefont{G\"{u}ntherodt}},
  \bibinfo{journal}{Solid State Commun.} \textbf{\bibinfo{volume}{35}},
  \bibinfo{pages}{1017} (\bibinfo{year}{1980}).

\bibitem[{\citenamefont{Werthamer et~al.}(1966)\citenamefont{Werthamer,
  Helfand, and Hohenberg}}]{WerthamerPR147}
\bibinfo{author}{\bibfnamefont{N.~R.} \bibnamefont{Werthamer}},
  \bibinfo{author}{\bibfnamefont{E.}~\bibnamefont{Helfand}}, \bibnamefont{and}
  \bibinfo{author}{\bibfnamefont{P.~C.} \bibnamefont{Hohenberg}},
  \bibinfo{journal}{Phys. Rev.} \textbf{\bibinfo{volume}{147}},
  \bibinfo{pages}{295} (\bibinfo{year}{1966}).

\bibitem[{\citenamefont{Poon}(1983)}]{Poon}
\bibinfo{author}{\bibfnamefont{S.~J.} \bibnamefont{Poon}},
  \bibinfo{journal}{Phys. Rev. B} \textbf{\bibinfo{volume}{27}},
  \bibinfo{pages}{5519} (\bibinfo{year}{1983}).

\bibitem[{\citenamefont{Domb and Johnson}(1978)}]{DombJLTP33}
\bibinfo{author}{\bibfnamefont{E.~R.} \bibnamefont{Domb}} \bibnamefont{and}
  \bibinfo{author}{\bibfnamefont{W.~L.} \bibnamefont{Johnson}},
  \bibinfo{journal}{J. Low Temp. Phys.} \textbf{\bibinfo{volume}{33}},
  \bibinfo{pages}{29} (\bibinfo{year}{1978}).

\bibitem[{\citenamefont{Gor'kov}(1959)}]{GorkovSPJETP9}
\bibinfo{author}{\bibfnamefont{L.~P.} \bibnamefont{Gor'kov}},
  \bibinfo{journal}{Sov. Phys. JETP} \textbf{\bibinfo{volume}{9}},
  \bibinfo{pages}{1364} (\bibinfo{year}{1959}).

\bibitem[{\citenamefont{Gor'kov}(1960)}]{GorkovSPJETP10b}
\bibinfo{author}{\bibfnamefont{L.~P.} \bibnamefont{Gor'kov}},
  \bibinfo{journal}{Sov. Phys. JETP} \textbf{\bibinfo{volume}{10}},
  \bibinfo{pages}{998} (\bibinfo{year}{1960}).

\bibitem[{\citenamefont{Kes and Tsuei}(1983)}]{KesPRB28}
\bibinfo{author}{\bibfnamefont{P.~H.} \bibnamefont{Kes}} \bibnamefont{and}
  \bibinfo{author}{\bibfnamefont{C.~C.} \bibnamefont{Tsuei}},
  \bibinfo{journal}{Phys. Rev. B} \textbf{\bibinfo{volume}{28}},
  \bibinfo{pages}{5126} (\bibinfo{year}{1983}).

\bibitem[{\citenamefont{McMillan}(1968)}]{McMillanPR167}
\bibinfo{author}{\bibfnamefont{W.~L.} \bibnamefont{McMillan}},
  \bibinfo{journal}{Phys. Rev.} \textbf{\bibinfo{volume}{167}},
  \bibinfo{pages}{331} (\bibinfo{year}{1968}).

\bibitem[{\citenamefont{Liu et~al.}(2002)\citenamefont{Liu, Luo, Leng, Wang,
  Qiu, Ding, and Lin}}]{LiuPRB66}
\bibinfo{author}{\bibfnamefont{Y.}~\bibnamefont{Liu}},
  \bibinfo{author}{\bibfnamefont{H.}~\bibnamefont{Luo}},
  \bibinfo{author}{\bibfnamefont{X.}~\bibnamefont{Leng}},
  \bibinfo{author}{\bibfnamefont{Z.~H.} \bibnamefont{Wang}},
  \bibinfo{author}{\bibfnamefont{L.}~\bibnamefont{Qiu}},
  \bibinfo{author}{\bibfnamefont{S.~Y.} \bibnamefont{Ding}}, \bibnamefont{and}
  \bibinfo{author}{\bibfnamefont{L.~Z.} \bibnamefont{Lin}},
  \bibinfo{journal}{Phys. Rev. B} \textbf{\bibinfo{volume}{66}},
  \bibinfo{pages}{144510} (\bibinfo{year}{2002}).

\bibitem[{\citenamefont{Xu et~al.}(2007)\citenamefont{Xu, Fangohr, Ding, Gu,
  Tang, Han, Shi, and Dou}}]{Xu}
\bibinfo{author}{\bibfnamefont{X.~B.} \bibnamefont{Xu}},
  \bibinfo{author}{\bibfnamefont{H.}~\bibnamefont{Fangohr}},
  \bibinfo{author}{\bibfnamefont{S.~Y.} \bibnamefont{Ding}},
  \bibinfo{author}{\bibfnamefont{M.}~\bibnamefont{Gu}},
  \bibinfo{author}{\bibfnamefont{T.~B.} \bibnamefont{Tang}},
  \bibinfo{author}{\bibfnamefont{Z.~H.} \bibnamefont{Han}},
  \bibinfo{author}{\bibfnamefont{D.~Q.} \bibnamefont{Shi}}, \bibnamefont{and}
  \bibinfo{author}{\bibfnamefont{S.~X.} \bibnamefont{Dou}},
  \bibinfo{journal}{Phys. Rev. B} \textbf{\bibinfo{volume}{75}},
  \bibinfo{eid}{224507} (\bibinfo{year}{2007}).

\bibitem[{\citenamefont{Ji et~al.}(1993)\citenamefont{Ji, Rzchowski, Anand, and
  Tinkham}}]{JiPRB47}
\bibinfo{author}{\bibfnamefont{L.}~\bibnamefont{Ji}},
  \bibinfo{author}{\bibfnamefont{M.~S.} \bibnamefont{Rzchowski}},
  \bibinfo{author}{\bibfnamefont{N.}~\bibnamefont{Anand}}, \bibnamefont{and}
  \bibinfo{author}{\bibfnamefont{M.}~\bibnamefont{Tinkham}},
  \bibinfo{journal}{Phys. Rev. B} \textbf{\bibinfo{volume}{47}},
  \bibinfo{pages}{470} (\bibinfo{year}{1993}).

\bibitem[{\citenamefont{Kili\c{c} et~al.}(2005)\citenamefont{Kili\c{c},
  Kili\c{c}, Yetis, and \c{C}etin}}]{Kilic}
\bibinfo{author}{\bibfnamefont{A.}~\bibnamefont{Kili\c{c}}},
  \bibinfo{author}{\bibfnamefont{K.}~\bibnamefont{Kili\c{c}}},
  \bibinfo{author}{\bibfnamefont{H.}~\bibnamefont{Yetis}}, \bibnamefont{and}
  \bibinfo{author}{\bibfnamefont{O.}~\bibnamefont{\c{C}etin}},
  \bibinfo{journal}{New Journal of Physics} \textbf{\bibinfo{volume}{7}},
  \bibinfo{pages}{212} (\bibinfo{year}{2005}).

\bibitem[{\citenamefont{dos Santos et~al.}(2003)\citenamefont{dos Santos,
  da~Luz, Ferreira, and Machado}}]{dosSantosPhysicaC391}
\bibinfo{author}{\bibfnamefont{C.~A.~M.} \bibnamefont{dos Santos}},
  \bibinfo{author}{\bibfnamefont{M.~S.} \bibnamefont{da~Luz}},
  \bibinfo{author}{\bibfnamefont{B.}~\bibnamefont{Ferreira}}, \bibnamefont{and}
  \bibinfo{author}{\bibfnamefont{A.~J.~S.} \bibnamefont{Machado}},
  \bibinfo{journal}{Physica C} \textbf{\bibinfo{volume}{391}},
  \bibinfo{pages}{345} (\bibinfo{year}{2003}).

\bibitem[{\citenamefont{Poon}(1982)}]{PoonPRB25}
\bibinfo{author}{\bibfnamefont{S.~J.} \bibnamefont{Poon}},
  \bibinfo{journal}{Phys. Rev. B} \textbf{\bibinfo{volume}{25}},
  \bibinfo{pages}{1977} (\bibinfo{year}{1982}).

\bibitem[{\citenamefont{Lefebvre}(2008)}]{myPhDthesis}
\bibinfo{author}{\bibfnamefont{J.}~\bibnamefont{Lefebvre}}, Ph.D. thesis,
  \bibinfo{address}{McGill University, {Montr\'{e}al}, Canada}
  (\bibinfo{year}{2008}).

\end{thebibliography}
\end{document}